\documentclass[aps,prb,onecolimn,preprintnumbers,amsmath,amssymb,superscriptaddress]{revtex4}%

\usepackage{amsmath,amsfonts}
\usepackage{graphicx}
\usepackage{hyperref}
\usepackage{float}
\usepackage[utf8x]{inputenc}
\usepackage{color}
\usepackage{multirow}

\newcommand{\RN}[1]{\textup{\uppercase\expandafter{\romannumeral#1}}}%

\begin{document}

\title{ Oxygen isotope effect on the superfluid density within the $d-$wave and $s-$wave pairing channels of YBa$_2$Cu$_4$O$_8$ }
\author{Rustem Khasanov}
 \email{rustem.khasanov@psi.ch}
 \affiliation{Laboratory for Muon Spin Spectroscopy, Paul Scherrer Institute, CH-5232 Villigen PSI, Switzerland}

\author{Alexander Shengelaya}
 \email{alexander.shengelaya@tsu.ge}
 \affiliation{Department of Physics, Tbilisi State University, Chavchavadze 3, GE-0128 Tbilisi, Georgia}
 \affiliation{Andronikashvili Institute of Physics of I.Javakhishvili Tbilisi State University, Tamarashvili  6, 0177 Tbilisi, Georgia}

\author{Kazimierz Conder}
 \affiliation{Laboratory for Multiscale Materials Experiments, Paul Scherrer Institut, CH-5232 Villigen, Switzerland}

\author{Janusz Karpinski}
 \affiliation{Laboratory for Solid State Physics, ETH Zurich, CH-8093 Zurich, Switzerland}

\author{Annette Bussmann-Holder}
 \email{A.Bussmann-Holder@fkf.mpg.de}
 \affiliation{Max-Planck-Institute for Solid State Research, Heisenbergstrasse 1, D-70569 Stuttgart, Germany}

\author{Hugo Keller}
 \email{keller@physik.uzh.ch}
 \affiliation{Physik-Institut der Universit\"{a}t Zürich, Winterthurerstrasse 190, CH-8057 Z\"{u}rich, Switzerland}

\begin{abstract}
We report on measurements of the oxygen isotope ($^{16}$O/$^{18}$O) effect (OIE) on the transition temperature $T_{\rm c}$ and the zero-temperature in-plane magnetic penetration depth $\lambda_{\rm ab}(0)$ in the stoichiometric cuprate superconductor YBa$_2$Cu$_4$O$_8$ by means of muon-spin rotation/relaxation.
An analysis of the temperature evolution of $\lambda^{-2}_{\rm ab}$ in terms of coexisting $s+d-$wave order parameters reveals that the OIE on
the superfluid density $\rho_{\rm s}(0)\propto\lambda^{-2}_{\rm ab}(0)$
stems predominantly from the $d-$wave component while the contribution of the $s-$wave one is almost zero.
The OIE on the transition temperature $T_{\rm c}$ is found to be rather small: $\delta T_{\rm c}/T_{\rm c}= -0.32(7)$\%, compared to the
total OIE on the superfluid density $\rho_{\rm s}(0)$: $\delta \rho_{\rm s}(0)/\rho_{\rm s}(0)= -2.8(1.0)$\%.
The partial OIE's on the corresponding $d-$wave and $s-$wave components of $\rho_{\rm s}(0)$ are $\delta \rho_{\rm s,d}(0)/\rho_{\rm s}(0)= -3.0(1.2)$\%, and $\delta \rho_{\rm s,s}(0)/\rho_{\rm s}(0)= 0.2(1.2)$\%, respectively.
Our results demonstrate that polaron formation in the CuO$_2$ planes is the origin of the observed OIE in the $d-$wave channel. In the much weaker $s-$wave channel, fermionic quasiparticles are present, which do not contribute to the OIE on $\rho_{\rm s}(0)$. Our results support the original idea of K. Alex M\"{u}ller on the polaronic nature of the supercarries in high-temperature cuprate superconductors.
\end{abstract}

\maketitle

\section{Introduction}
%
Almost 4 decades after the breakthrough discovery of high-temperature superconductivity in cuprates by Bednorz and M\"{u}ller \cite{Bednorz_Z-Phys-B_1986} the initial hype and intensive research activities have slightly faded in spite of the fact that essential questions remain unanswered. This holds especially for the pairing glue and, intimately related to it, the symmetry of the order parameter.
Even though the discovery of high-temperature superconductivity was inspired by polaron formation in connection with the Jahn-Teller (JT) effect, a consensus on its role could not be achieved. However, rapidly after the discovery, there was a certain agreement that the order parameter is of $d-$wave symmetry, thus favoring some purely electronic or magnetic pairing mechanisms. Yet, many experimental results did not support such statment and the lacking agreement between expectations and the data pushed K. Alex M\"{u}ller to the conclusion that a more complex answer is needed.
His careful inspection of the existing controversial results bounced him to the implication that coexisting $s+d-$wave symmetries are realized in cuprate high-temperature superconductors (HTSs).\cite{Muller_Nature_1995, Muller_Kluver_1997, Muller_PhilMagLett_2002,Muller_JPCM_2007}  While the CuO$_2$ planes are almost in accordance with a $d-$wave order parameter, along the $c-$axis an $s-$wave order parameter is realized.
Since the existence of multiple order parameters requires interactions between them, each individual component carries to a certain amount features from the others. In addition, his starting point, the JT polaron (a strong coupling electron-lattice interacting object) is intimately related to isotopic effects, can be conclusively supported and proven by the investigation of isotope effects on fundamental properties of HTSs. \cite{Muller_JPCM_2007,Muller_JSNM_2017} Consequently, he founded in 1990 a research group at the Physik-Institut of the University of Zurich dedicated exclusively to this topic, which very successfully discovered numerous unexpected isotope effects in HTSs and, thus, a strong support of polaron formation in HTSs (see, {\it e.g.}, Refs.~\onlinecite{Keller_OIE_Springer_2005, Keller_MaterToday_2008} and references therein).
Both of the above mentioned basic properties of HTSs are addressed in the following, namely coexisting $s+d-$wave order parameters and the oxygen isotope effect with emphasis on the magnetic field penetration depth ($\lambda$), a fundamental property of superconductors. Such studies have already been performed and published earlier, and amazing novel results could be achieved (see, {\it e.g.}, Ref.~\onlinecite{Keller_MaterToday_2008}). Especially, it was observed that $\lambda$ is strongly anisotropic where data along the $a-$axis are different from $b-$ and $c-$axis and vice versa. The most prominent differences are between planar and $c-$axis data where $d-$wave is mainly seen in the planes, whereas $s-$wave is dominant perpendicular to them. \cite{Khasanov_La214_PRL_2007,Khasanov_Y123_PRL_2007,Khasanov_Y124_JSNM_2008}

Experimentally, $\lambda$ is best investigated by muon-spin rotation ($\mu$SR) which is a direct and bulk testing technique.\cite{Blundell_book_2022, Yaouanc_book_2011}
It has already been used in earlier studies on different cuprates by measuring the temperature dependence of the magnetic penetration depth in the $ab-$plane (La$_{1.83}$Sr$_{0.17}$CuO$_{4}$ \cite{Khasanov_La214_PRL_2007}), and along the cystallographic $a$, $b$, and $c-$axes (YBa$_{2}$Cu$_{3}$O$_{7-\delta}$ \cite{Khasanov_Y123_PRL_2007} and YBa$_{2}$Cu$_{4}$O$_{8}$ \cite{Khasanov_Y124_JSNM_2008}) with the universal result that the in-plane order parameter (superfluid density) has a dominant $d-$wave component ($\approx$ 80\%) and a much smaller $s-$wave component ($\approx$ 20\%), whereas along the $c-$axis the order parameter has almost pure $s-$wave character, in accordance with the prediction of K. Alex M\"{u}ller.\cite{Muller_Nature_1995,Muller_Kluver_1997,Muller_PhilMagLett_2002,Muller_JPCM_2007}

In this work a detailed study of the oxygen isotope ($^{16}$O/$^{18}$O) effect (OIE) on the in-plane magnetic penetration depth ($\lambda_{ab}$) of the stochiometric HTS YBa$_2$Cu$_4$O$_8$ is presented. In particular, we focus on the OIE on the $d-$ and $s-$wave components of the superfluid density $\rho_{\rm s}(T) \propto \lambda_{\rm ab}^{-2}(T)$ by analysing carefully previous and unpublished $\mu$SR data of YBa$_2$Cu$_4$O$_8$.\cite{Khasanov_OIE_HTSs_lambda_PRB_2007, comment} It is found that the OIE on $\rho_{\rm s}\propto \lambda_{\rm ab}^{-2}$ arises predominantly from the $d-$wave component, whereas the $s-$wave channel exhibits no measurable OIE. It is important to highlight that in the present $\mu$SR study bulk properties are measured. This is in contrast to many other experimental techniques where only surface sensitive properties are obtained. Since the CuO$_2$ planes have been identified with the $d-$wave channel, features from the $c-$axis have been overlooked, simply due to the experimental constrains. These are, however, related to the $s-$wave order parameter which consequently could not be observed by these specific experimental tools.

\section{Experimental Details} \label{seq:Experimental Details}

The synthesis procedure of the polycrystalline YBa$_2$Cu$_4$O$_8$ sample is described elsewhere. \cite{Karpinsky_PhysicaC_1989, Bucher_PhysicaC_1989}
Oxygen isotope exchange was carried through by annealing the sample in $^{18}$O$_2$ gas. In order to ensure that the
$^{16}$O and $^{18}$O substituted samples are subjected to the same thermal history, the annealing of the pair of samples was performed simultaneously in $^{16}$O$_2$ and $^{18}$O$_2$ (95\% enriched) atmospheres. The amount of $^{18}$O substitution was estimated to be $\simeq 82$\%.\cite{Khasanov_OIE_HTSs_lambda_PRB_2007}

Prior to the $\mu$SR experiments, the $^{16}$O and $^{18}$O  YBa$_2$Cu$_4$O$_8$ samples were cold pressed into pellets ($\simeq10$~mm in diameter and $\simeq 1$~mm thick) and mounted back-to-back onto a fork-type sample holder.\cite{Amato_RSI_2017} Considering the stopping range of the surface muons used in our study to be $\simeq 0.3-0.5$~mm (see, {\it e.g.}, Fig. 2 in Ref.~\onlinecite{Khasanov_JAP_2022}), this setup allowed (i) for an easy access of the muon beam to the particular sample via rotation of the sample stick between 0 and 180 degrees and (ii) to ensure the same experimental conditions for both the $^{16}$O and $^{18}$O YBa$_2$Cu$_4$O$_8$ samples.

The transverse-field (TF) $\mu$SR experiments were carried out at the $\pi$M3 beam line by using the GPS (General Purose Surface) $\mu$SR spectrometer (Paul Scherrer Institut, Villigen, Switzerland).\cite{Amato_RSI_2017}  Experiments were conducted in the temperature range 1.7 -- 100~K in an applied magnetic field of $B_{\rm ap}=0.2$~T. The TF-$\mu$SR spectra were collected upon cooling and warming the samples in a constant applied magnetic field and were analyzed by using the Musrfit package.\cite{MUSRFIT}
Part of the raw $\mu$SR data were initially  presented in Ref.~\onlinecite{Khasanov_OIE_HTSs_lambda_PRB_2007}. However, the aspects highlighted here, namely the coexistence of $s+d-$wave order parameters, was overlooked before and the data were reanalyzed with respect to this specific feature.\cite{comment}

\section{Results}\label{seq:Results_and_discussion}

\begin{figure}[htb]
\includegraphics[width=0.8\textwidth]{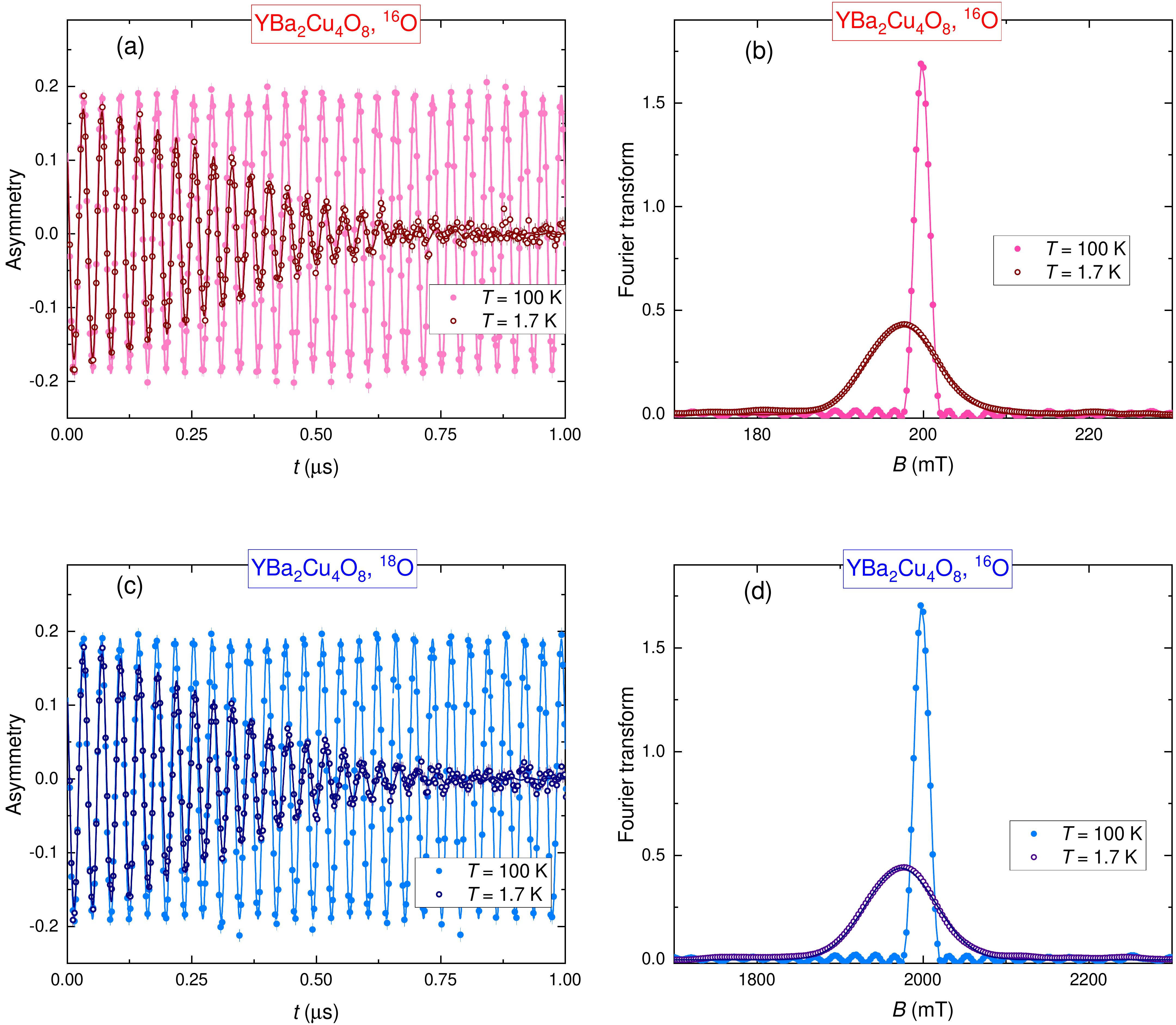}
\caption{ (a) $\mu$SR time spectra of the $^{16}$O YBa$_2$Cu$_4$O$_8$ sample collected above ($T=100$~K) and below ($T=1.7$~K) the superconducting transition temperature ($T_{\rm c}\simeq 80$~K) at an applied field of $B_{\rm ap}=0.2$~T. (b) Fourier transforms of the $\mu$SR time spectra shown in panel (a). The solid lines are fits of Eqs.~(\ref{eq:P(t)}) and (\ref{eq:P(B)}) to the $\mu$SR data (see text for details). (c) and (d) correspond to panels (a) and (b) but for the $^{18}$O substituted YBa$_2$Cu$_4$O$_8$ sample.}
\label{fig:time-spectra_fourier}
\end{figure}

Figures~\ref{fig:time-spectra_fourier}~(a) and (c) show the $\mu$SR time spectra of the $^{16}$O and $^{18}$O YBa$_2$Cu$_4$O$_8$ samples collected above ($T\simeq 100$~K) and below ($T\simeq 1.7$~K) the superconducting transition temperature ($T_{\rm c}\simeq 80$~K) at an applied magnetic field of $B_{\rm ap}=0.2$~T.
A strong damping at $T<T_{\rm c}$ reflects the inhomogeneous field distribution $P(B)$ caused by the formation of the flux-line lattice (FLL) in the mixed state. The broadening and the shift of the $P(B)$ distributions in the superconducting state to lower field values are clearly visible in Figs.~\ref{fig:time-spectra_fourier}~(b) and (d), where also the Fourier transforms  of the corresponding TF-$\mu$SR time spectra are shown.

Figures~\ref{fig:time-spectra_fourier}~(b) and (d) show that the shapes of $P(B)$ at $T=1.7$~K are symmetric and well described by a single Gaussian curve. This is an indication that the anisotropy ratio $\gamma=\lambda_{\rm c}/\lambda_{\rm ab}$ is large for YBa$_2$Cu$_4$O$_8$ ($\lambda_{\rm c}$ and $\lambda_{\rm ab}$ are the out-of plane and the in-plane components of the magnetic penetration depth, respectively).\cite{Khasanov_Y124_JSNM_2008} Following Ref~\onlinecite{Pumpin_PRB_1990}, the field distribution in the sintered superconducting sample with a large $\gamma$ is mainly influenced by the shortest component of the magnetic penetration depth $\lambda_{\rm ab}$ and is almost independent of $\lambda_{\rm c}$.

\begin{figure}[htb]
\includegraphics[width=0.5\textwidth]{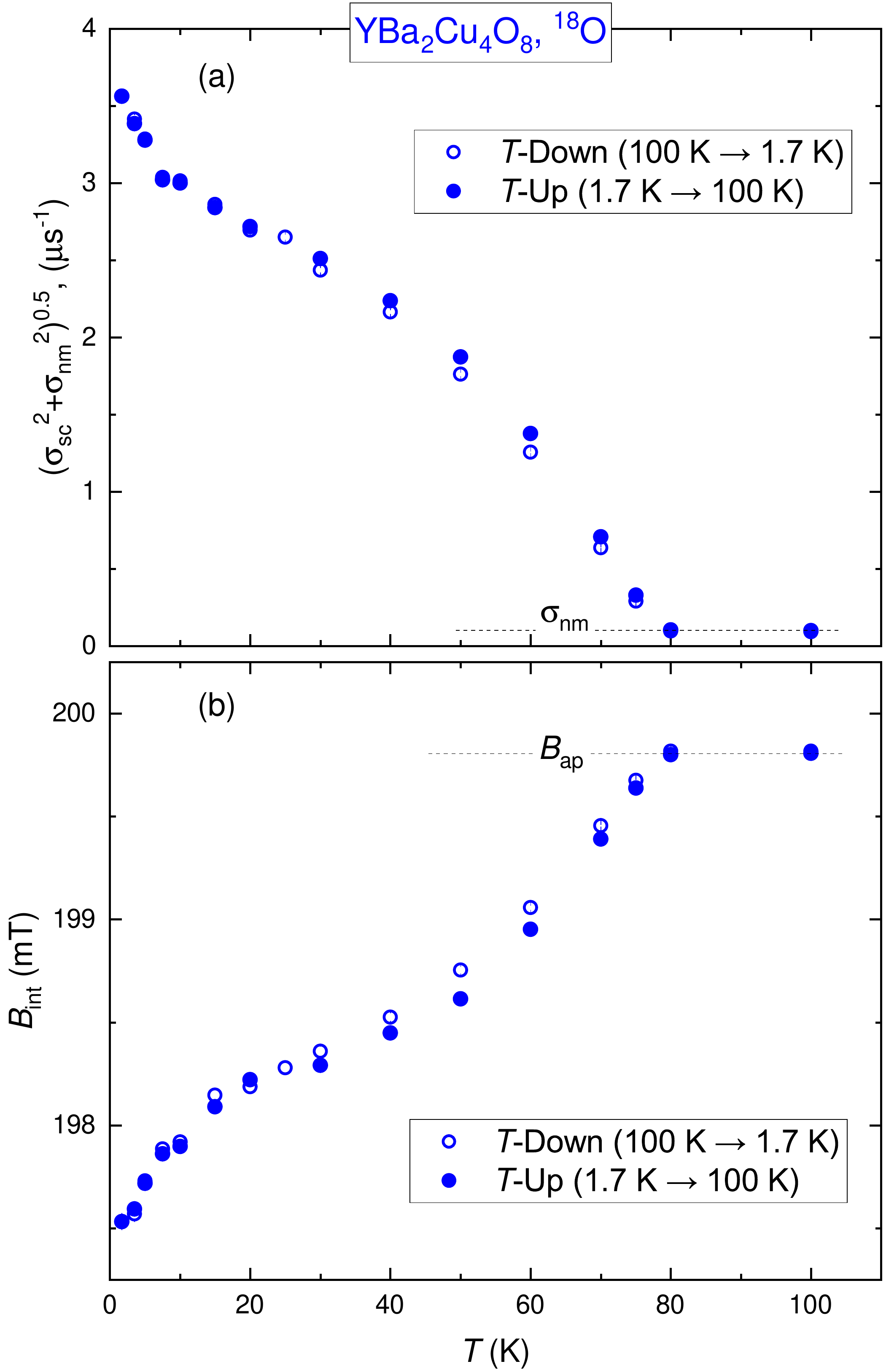}
\caption{ (a) Temperature dependence of the combined Gaussian depolarization rate $\sigma=(\sigma_{\rm sc}^2+\sigma_{\rm nm}^2)^{0.5}$ of the $^{18}$O substituted YBa$_2$Cu$_4$O$_8$ sample. (b) Temperature dependence of the internal field $B_{\rm int}$ of $^{18}$O YBa$_2$Cu$_4$O$_8$ sample. The open and closed symbols correspond to the data accumulated by following the $T-$Down and $T-$Up measurement protocols, respectively (see text for details).}
 \label{fig:Sigma_Bint_18O}
\end{figure}

The TF-$\mu$SR spectra were analyzed by using a simple Gaussian function for the time evolution of the muon-spin polarization $P(t)$:
\begin{equation}
A(t) = A_0  P(t) = A_{0}\; \exp\left[ { -\frac{(\sigma_{\rm sc}^2+\sigma_{\rm nm}^2)t^2}{2}}\right]\; \cos (\gamma_\mu B_{\rm int} t +\phi).
\label{eq:P(t)}
\end{equation}
Here $A_0$ is the initial asymmetry of the muon-spin polarization, $\sigma_{\rm sc}$ is the Gaussian relaxation rate caused by the formation of a FLL in the superconducting state, $\sigma_{\rm nm}$ is the temperature independent nuclear moment contribution,  $\gamma_\mu = 2\pi\times135.5342$~MHz/T is the muon gyromagnetic ratio, $B_{\rm int}$ is the internal field at the muon-stopping site, and $\phi$ is the initial phase of the muon-spin ensemble. Note that for $T > T_{\rm c}$ the internal field is equal to the applied field $B_{\rm int}=B_{\rm ap}$ and only the nuclear moment contribution $\sigma=\sigma_{\rm nm}$ is present [see also the dashed lines in Figs.~\ref{fig:Sigma_Bint_18O}~(a) and (b)].
$P(t)$ given in Eq.~(\ref{eq:P(t)}) corresponds to the field distribution $P(B)$:
\begin{equation}
P(B)=\frac{\gamma_\mu A_{0}}{\sqrt{\sigma_{\rm sc}^2+\sigma_{\rm nm}^2}} \exp \left(-\frac{\gamma_\mu^2(B-B_{\rm int})^2}{2(\sigma_{\rm sc}^2+\sigma_{\rm nm}^2)} \right)  %
\label{eq:P(B)}
\end{equation}
The solid lines in Fig.~\ref{fig:time-spectra_fourier} represent fits to the $\mu$SR data using Eqs.~(\ref{eq:P(t)}) and (\ref{eq:P(B)}).

\begin{figure}[htb]
\includegraphics[width=0.8\textwidth]{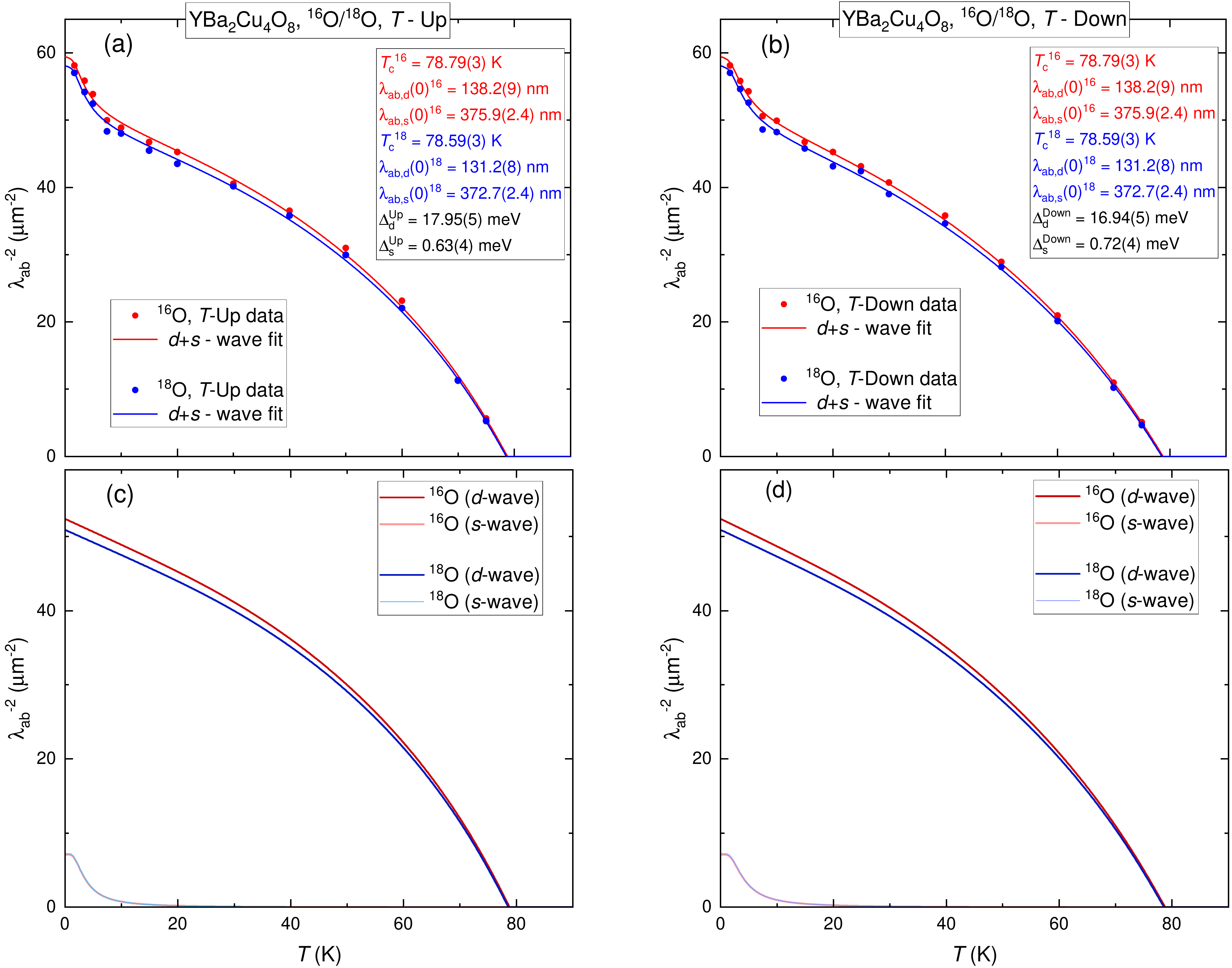}
\caption{ (a) The temperature dependencies of the inverse squared in-plane magnetic penetration depth $\lambda_{\rm ab}^{-2}$ for the $^{16}$O and $^{18}$O substituted YBa$_2$Cu$_4$O$_8$ samples, as obtained by following the '$T-$Up' protocol. The solid lines are fits of Eq.~(\ref{eq:two-gap}) to the data (see text for details). (b) The same as in (a), but for the data collected by following the '$T-$Down' protocol. (c) and (d) The individual $d-$wave and $s-$wave components of $\lambda_{\rm ab}^{-2}(T)$   derived by the two-gap fit described in the text.}
 \label{fig:Lambda_16O-18O}
\end{figure}

It should be emphasized here, that experiments on the isotope substituted samples should be undertaken under {\it exactly the same} experimental conditions. As an example, Fig.~\ref{fig:Sigma_Bint_18O} shows the temperature dependencies of two relevant fit parameters: the combined Gaussian relaxation rate $\sigma=(\sigma_{\rm sc}^2+\sigma_{\rm nm}^2)^{0.5}$ and the internal field $B_{\rm int}$, obtained for the $^{18}$O substituted YBa$_2$Cu$_4$O$_8$ sample for two different measurement sets. The first set of experiments, denoted as '$T-$ Down', corresponds to a cooling cycle (from $T\simeq 100$~K to $T=1.7$~K) with the temperature stabilization and data accumulation at temperature steps of 5 to 10~K. In the second set of experiments, '$T-$Up',  the sample was warmed up (from $T=1.7$~K to $T\simeq 100$~K) by following almost identical  temperature steps. Figure~\ref{fig:Sigma_Bint_18O} implies that the low-temperature ($1.7\lesssim T \lesssim 20$~K) and the high-temperature ($75\lesssim T \lesssim 100$~K) values of $\sigma$ and $B_{\rm int}$, respectively, coincide with each other, while tiny differences are obvious in the intermediate temperature range ($25\lesssim T \lesssim 70$~K). This could be caused by slightly different arrangement of the FLL during the cooling and warming processes. Note that in the reminder of the paper we are comparing experimental data collected under similar thermal histories.

From the measured $\mu$SR relaxation rate $\sigma=(\sigma_{\rm sc}^2+\sigma_{\rm nm}^2)^{0.5}$ determined below $T_{c}$ and the nuclear relaxation rate $\sigma_{\rm nm}$ determined above $T_{c}$ the
absolute value of the in-plane magnetic penetration depth $\lambda_{\rm ab}$ was determined by using the relation: $\sigma_{\rm sc}^2 /\gamma_{\mu}^2 = 0.00126 \; \Phi_0^2/\lambda_{\rm ab}^{-4} $, $\Phi_0=2.026\times10^{-15}$~Wb is the magnetic flux quantum,
yielding the practical relation to determine $\lambda_{\rm ab}$ from the measured $\sigma_{\rm sc}$: $\lambda_{\rm ab}[{\rm nm}] \simeq 250/\sqrt{\sigma_{\rm sc} [{\mu \rm s^{-1}}]}$.
\cite{Brandt_PRB_1988, Fesenko_PhysicaC_1991, Khasanov_Sm1111-Nd1111_PRB_2008}\
A physical quantity of special interest in this work is the superfluid density $\rho_{\rm s}$ which is related to $\lambda_{\rm ab}$ (see, {\it e.g.},  Ref.~\onlinecite{Keller_OIE_Springer_2005}):
\begin{equation}
\rho_{\rm s} (T) \propto \sigma_{\rm sc}(T) \propto \lambda^{-2}_{\rm ab}(T) \propto n_{\rm s}/m^{\ast}_{\rm ab},
\label{eq:rho_s}
\end{equation}
where $n_{\rm s}$ is the superconducting carrier density, and $m^{\ast}_{\rm ab}$ is the in-plane effective mass of the carriers.

Figure~\ref{fig:Lambda_16O-18O} compares the temperature dependencies of $\lambda^{-2}_{\rm ab}$ for the $^{16}$O and $^{18}$O substituted YBa$_2$Cu$_4$O$_8$ samples. Panels (a) and (b) represent the data collected during the '$T-$Up' and '$T-$Down' cycles, respectively.
From the data presented in Figs.~\ref{fig:Lambda_16O-18O}~(a) and (b) the following two important points arise:
(i) For both measurement protocols ('$T-$Up' and '$T-$Down'), the values of $\lambda^{-2}_{\rm ab}$ for the $^{18}$O substituted YBa$_2$Cu$_4$O$_8$ sample are systematically lower as compared to those for the $^{16}$O sample. This is in agreement with the wide variety of data demonstrating the existence of an oxygen isotope effect (OIE) on $T_{\rm c}$ and $\lambda_{ab}$ in various families of HTSs.\cite{Zhao_Nature_1997, Zhao_JPCM_1998, Hofer_PRL_2000, Khasanov_YPr123_JPCM_2003, Khasanov_SSOIE-YPr123_PRB_2003, Khasanov_LEM-OIE_PRL_2004, Khasanov_Y123-La214_PRB_2006}
(ii) For both samples the temperature dependence of $\lambda_{\rm ab}^{-2}$ shows an inflection point at $T\simeq 10$~K, which is a clear signature of the presence of two superconducting energy gaps with largely different zero-temperature values.\cite{Khasanov_La214_PRL_2007, Khasanov_Y124_JSNM_2008, Khasanov_Y123_PRL_2007, Khasanov_SFCA_PRL_2009}

The temperature dependence of $\lambda_{\rm ab}^{-2}$ was further analyzed within the framework of the simplified phenomenological $\alpha-$model by assuming the presence of a $d-$ and an $s-$wave component of the order parameter without coupling between them using the relation:\cite{Khasanov_La214_PRL_2007}

\begin{equation}
\lambda_{\rm ab}^{-2}(T) = \lambda_{\rm ab,d}^{-2}(T)+\lambda_{\rm ab,s}^{-2}(T)
\label{eq:two-gap}
\end{equation}
Here indices 'd' and 's' refer to the $d-$wave and the $s-$wave contribution, respectively.
Each individual component at the right-hand site of Eq.~(\ref{eq:two-gap}) was calculated within the local (London) approach by using the following functional form: \cite{Tinkham_75, Khasanov_La214_PRL_2007}
\begin{equation}
\frac{\lambda_{\rm ab,d(s)}^{-2}(T)}{\lambda_{\rm ab,d(s)}^{-2}(0)}=  1
+\frac{1}{\pi}\int_{0}^{2\pi}\int_{\Delta_{\rm d(s)}(T,\varphi)}^{\infty}\left(\frac{\partial
f}{\partial E}\right)\frac{E\
dEd\varphi}{\sqrt{E^2-\Delta_{\rm d(s)}(T,\varphi)^2}}~.
 \label{eq:lambda-d}
\end{equation}
Here $\lambda_{\rm ab, d(s)}^{-2}(0)$ is the zero-temperature value of the $d-$wave ($s-$wave) component defined in Eq.~(\ref{eq:two-gap}), $f=[1+\exp(E/k_BT)]^{-1}$ is  the Fermi function, $\varphi$ is the angle along the Fermi surface,
and $\Delta_{\rm d(s)}(T,\varphi)=\Delta_{\rm d(s)} \ g_{\rm d(s)}(\varphi) \tanh\{1.82[1.018(T_c/T-1)]^{0.51}\}$ is the temperature dependent superconducting energy gap.\cite{Khasanov_La214_PRL_2007} $g_{\rm d(s)}(\varphi)$ describes the angular dependence of the gap: $g_d(\varphi)=\cos(2\varphi)$ for the $d-$wave gap, and $g_s(\varphi)=1$ for the $s-$wave gap; and $\Delta_{\rm d(s)}$ is the zero-temperature gap value.

It should be noted that a fit of Eqs.~(\ref{eq:two-gap}) and (\ref{eq:lambda-d}) to a single $\lambda_{\rm ab}^{-2}(T)$ data set requires the use of 5 independent parameters, namely $T_{\rm c}$, $\lambda_{\rm ab,d}(0)$, $\lambda_{\rm ab,s}(0)$, $\Delta_{\rm d}$, and $\Delta_{\rm s}$. With 4 data sets (2 for each $^{16}$O and $^{18}$O substituted YBa$_2$Cu$_4$O$_8$ samples) the total number of independent parameters reaches 20. Obviously, not all these parameters are independent, and some of them may stay the same for different data sets. More important, the reduced number of the parameters may improve the fit quality. We have performed a simultaneous fit of all 4 $\lambda_{\rm ab}^{-2}(T)$ data sets by searching for 'global' parameters via minimization of the reduced $\chi^2_r$ (the sum of mean square deviations divided by the number of degrees of freedom), {\it i.e.} via controlling the 'goodness' of the fit. The best result (minimum of $\chi^2_r$) was obtained by considering similar values of $T_{\rm c}$, $\lambda_{\rm ab,d}(0)$, and $\lambda_{\rm ab,s}(0)$ for the '$T-$Up'/'$T-$Down' scans of each isotope enriched sample [$T_{\rm c}^{16}$, $\lambda_{\rm ab,d}(0)^{16}$, $\lambda_{\rm ab,s}(0)^{16}$; and $T_{\rm c}^{18}$, $\lambda_{\rm ab,d}(0)^{18}$, $\lambda_{\rm ab,s}(0)^{18}$], and assuming  the superconducting energy gaps to be the same for the '$T-$Up' and '$T-$Down' scans ($\Delta_{\rm d}^{\rm Up}$, $\Delta_{\rm s}^{\rm Up}$ and $\Delta_{\rm d}^{\rm Down}$, $\Delta_{\rm s}^{\rm Down}$). Such a "parameter-restricted fit" allows to reduce the number of parameters from 20 down to 10  and to reduce the normalized $\chi^2_r$ from $\chi^2_r\simeq4.5$ in the case of 20 parameters down to $\chi^2_r\simeq3.7$ for 10 parameters. The results of this procedure are summarized in Table~\ref{tab:Fit-results}. The $s+d-$wave fitting curves are represented by the solid lines in Figs.~\ref{fig:Lambda_16O-18O} (a) and (b). The corresponding individual $d-$wave and $s-$wave components are shown in Figs.~\ref{fig:Lambda_16O-18O}~(c) and (d).

\begin{table}[htb]
     \centering
     \caption{The results of the fit of Eqs.~(\ref{eq:two-gap}) and (\ref{eq:lambda-d}) to the $^{16}$O/$^{18}$O $\lambda_{\rm ab}^{-2}(T)$ data of YBa$_2$Cu$_4$O$_8$. $T_{\rm c}$ is the superconducting transition temperature, $\lambda_{\rm ab}(0)$  is the zero-temperature in-plane magnetic penetration depth, and $\lambda_{\rm ab,d(s)}(0)$ and $\Delta_{\rm d(s)}$ are the zero-temperature values of the $d-$wave($s-$wave) components related to the superfluid density ($\lambda_{\rm ab}^{-2}$) and the energy gap, respectively. \\}
     \begin{tabular}{l|cccccc}
 \hline  \hline
Sample/Protocol & $T_{\rm c}$ & $\lambda_{\rm ab}(0)$ &  $\lambda_{\rm ab,d}(0)$ & $\lambda_{\rm ab,s}(0)$ & $\Delta_{\rm d}$ & $\Delta_{\rm s}$ \\
                &  (K)        &       (nm)          &   (nm)    &   (nm)    &        (meV)     &   (mev)           \\
 \hline
$^{16}$O / $T-$Up & \multirow{2}{*}{78.79(3)} & \multirow{2}{*}{129.7(8)} & \multirow{2}{*}{138.2(9)} & \multirow{2}{*}{375.9(2.4)} & 17.95(5)& 0.63(4) \\
$^{16}$O / $T-$Down & & & & & 16.94(5) & 0.72 (4) \\
 \hline
$^{18}$O / $T-$Up & \multirow{2}{*}{78.59(3)} & \multirow{2}{*}{131.2(8)} & \multirow{2}{*}{140.2(9)} & \multirow{2}{*}{372.7(2.3)} &\multicolumn{2}{l}{The same as for $^{16}$O / $T-$Up}   \\
$^{18}$O / $T-$Down & & & & & \multicolumn{2}{l}{The same as for $^{16}$O / $T-$Down}   \\
 \hline  \hline
     \end{tabular}
     \label{tab:Fit-results}
 \end{table}
%

The results presented in Table~\ref{tab:Fit-results} allow to determine the OIE shifts of $T_{\rm c}$ and $\rho_{\rm s}(0) \propto\lambda_{\rm ab}^{-2}(0)$:
\begin{equation}
\frac{\delta T_{\rm c}}{T_{\rm c}}= \frac{T_{\rm c}^{18}-T_{\rm c}^{16}}{T_{\rm c}^{16}},
\label{eq:OIE-Tc}
\end{equation}
\begin{equation}
\frac{\delta\rho_{\rm s}(0)}{\rho_{\rm s}(0)}\equiv\frac{\delta \lambda_{\rm ab}^{-2}(0)}{\lambda_{\rm ab}^{-2}(0)}= \frac{\lambda_{\rm ab}^{-2}(0)^{18}- \lambda_{\rm ab}^{-2}(0)^{16}}{\lambda_{\rm ab}^{-2}(0)^{16}}.
 \label{eq:OIE-lambda-2_total}
\end{equation}
By using Eq.~(\ref{eq:two-gap}) one obtains for the partial OIE shifts of the $d-$wave and $s-$wave components of the superfluid density:
\begin{equation}
\frac{\delta\rho_{\rm s,d(s)}(0)}{\rho_{\rm s}(0)}\equiv\frac{\delta \lambda_{\rm ab,d(s)}^{-2}(0)}{\lambda_{\rm ab}^{-2}(0)}= \frac{\lambda_{\rm ab,d(s)}^{-2}(0)^{18}- \lambda_{\rm ab,d(s)}^{-2}(0)^{16}}{ \lambda_{\rm ab}^{-2}(0)^{16}}.
 \label{eq:OIE-lambda-2_d-channel}
\end{equation}
Note that Eqs.~(\ref{eq:OIE-lambda-2_total}) and (\ref{eq:OIE-lambda-2_d-channel}) are not independent of each other, but they are related via:
\begin{equation}
\frac{\delta \lambda_{\rm ab}^{-2}(0)}{\lambda_{\rm ab}^{-2}(0)}= \frac{\delta \lambda_{\rm ab,d}^{-2}(0)}{\lambda_{\rm ab}^{-2}(0)}+\frac{\delta \lambda_{\rm ab,s}^{-2}(0)}{\lambda_{\rm ab}^{-2}(0)}. \nonumber
\end{equation}

By taking an $^{18}$O$_2$ enrichment of $\simeq 82$\% in the $^{18}$O substituted YBa$_2$Cu$_4$O$_8$ sample into account,\cite{Khasanov_OIE_HTSs_lambda_PRB_2007} and using the relations above with the corresponding values listed in Table~\ref{tab:Fit-results} one obtains the following OIE shifts:
\begin{center}
$\delta T_{\rm c}/T_{\rm c}= -0.32(7)$\%,
\end{center}
and
\begin{center}
$\delta \lambda_{\rm ab}^{-2}(0)/\lambda_{\rm ab}^{-2}(0)=-2.8(1.0)$\%, $\delta \lambda_{\rm ab,d}^{-2}(0)/\lambda_{\rm ab,d}^{-2}(0)= -3.0(1.2)$\%, and $\delta \lambda_{\rm ab,s}^{-2}(0)/\lambda_{\rm ab,s}^{-2}(0)=0.2(1.2)$\%.
\end{center}
This implies that the total OIE on the superfluid density $\rho_{\rm s}(0)\propto\lambda^{-2}_{\rm ab}(0)$ is almost entirely determined by the $d-$wave channel, while the OIE on
$\lambda^{-2}_{\rm ab}(0)$ for the $s-$wave part is zero within the experimental error: $\delta \lambda^{-2}_{\rm ab}(0)/\lambda^{-2}_{\rm ab}(0) \simeq \,\, \delta \lambda_{\rm ab,d}^{-2}(0)/\lambda_{\rm ab}(0)^{-2} \simeq -3\%$.
\begin{figure}[htb]
\includegraphics[width=0.6\textwidth]{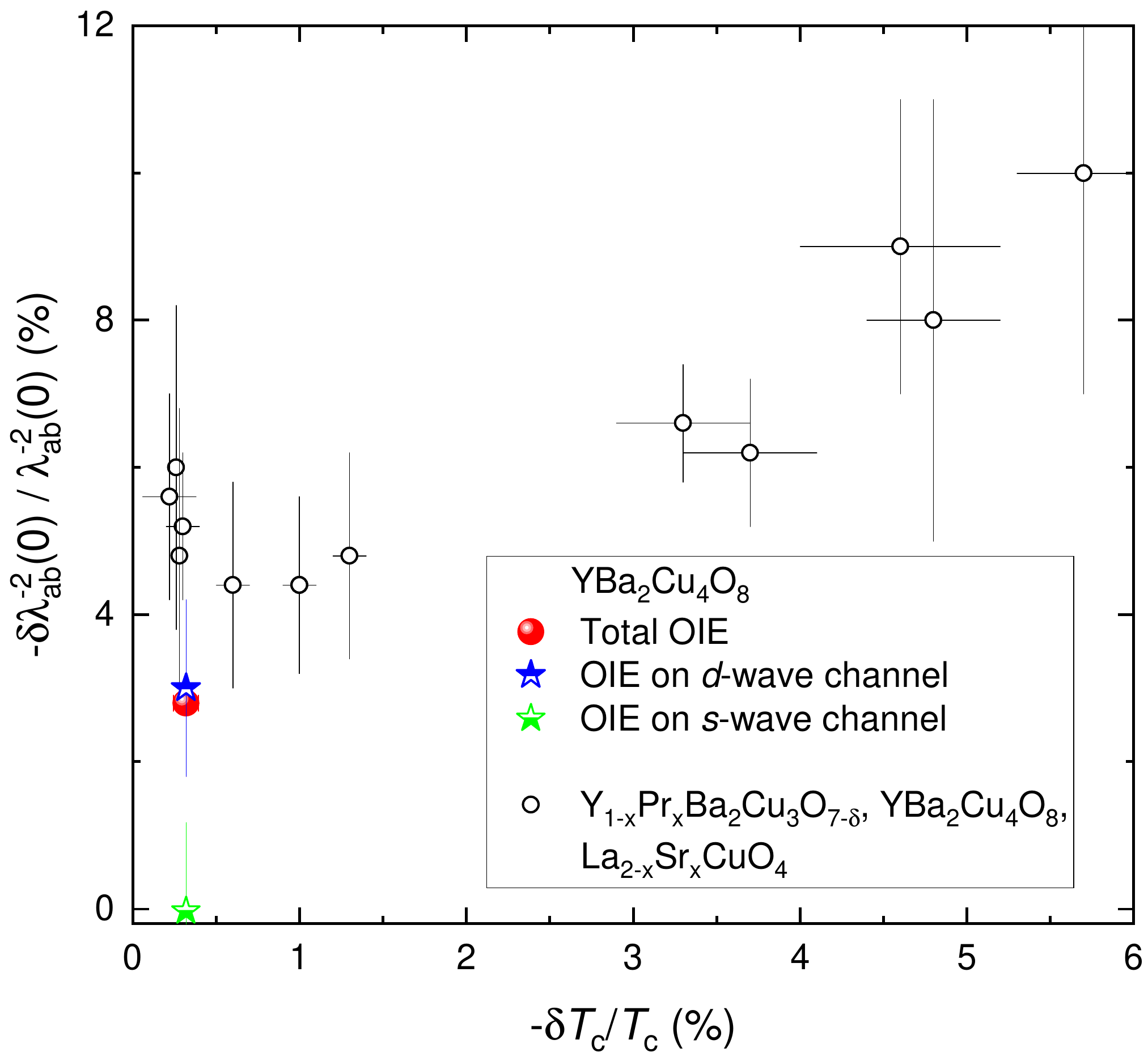}
\caption{The OIE shift of the zero-temperature superfluid density $\delta\rho_{\rm s}(0)/\rho_{\rm s}(0)\equiv\delta \lambda^{-2}_{ab}(0)/\lambda^{-2}_{\rm ab}(0)$ versus the OIE shift of the superconducting transition temperature $\delta T_{\rm c}/T_{\rm c}$ for various families of HTSs. The colored symbols represent the results of the present study. The black open symbols are data from Refs.~\onlinecite{Khasanov_OIE_HTSs_lambda_PRB_2007, Zhao_Nature_1997, Zhao_JPCM_1998, Hofer_PRL_2000, Khasanov_YPr123_JPCM_2003, Khasanov_SSOIE-YPr123_PRB_2003, Khasanov_LEM-OIE_PRL_2004, Khasanov_Y123-La214_PRB_2006}.}
 \label{fig:OIE_on_Tc_and_lambda}
\end{figure}

Figure~\ref{fig:OIE_on_Tc_and_lambda} compares the present results with those of previous OIE studies on $T_{\rm c}$ and $\lambda^{-2}_{\rm ab}(0)$ for
YBa$_2$Cu$_4$O$_8$ \cite{Khasanov_OIE_HTSs_lambda_PRB_2007} and other families of HTSs. \cite{Zhao_Nature_1997, Zhao_JPCM_1998, Hofer_PRL_2000, Khasanov_YPr123_JPCM_2003, Khasanov_SSOIE-YPr123_PRB_2003, Khasanov_LEM-OIE_PRL_2004, Khasanov_Y123-La214_PRB_2006}
Obviously, the OIE's on $T_{\rm c}$, $\lambda^{-2}_{\rm ab}(0)$ and $\lambda^{-2}_{\rm ab,d}(0)$ of YBa$_2$Cu$_4$O$_8$ are consistent with the general trend observed for various cuprate families. By considering the presence of two type of carriers with $d-$wave and $s-$wave symmetry, we may conclude that the OIE on $\lambda_{\rm ab}^{-2}(0)$ in YBa$_2$Cu$_4$O$_8$ is associated with the $d-$wave type carriers, while the OIE on the $s-$wave type of carriers is zero.

\section{Discussion and Conclusions}

Site-selective oxygen isotope effect studies of Pr-doped   Y$_{1-x}$Pr$_x$Cu$_3$O$_{7-\delta}$ demonstrated that the OIE on $T_{\rm c}$ and the in-plane magnetic penetration depth $\lambda_{\rm ab}(0)$ are mainly due to the oxygen atoms in the CuO$_2$ planes (so-called planar oxygen atoms) and not by the apical or chain oxygen atoms.\cite{Khasanov_SSOIE-YPr123_PRB_2003, Zhao_PRB_1996} The present experiments show that the $d-$wave and the $s-$wave components of $\lambda_{\rm ab}^{-2}$ (superfluid density) exhibit an analogous behavior.
Indeed, the $d-$wave component shows the dominant contribution to the OIE on $\lambda_{\rm ab}^{-2}(0)$ and, correspondingly we associate it with planar oxygens.
The $s-$wave component shows no significant contribution to the OIE on $\lambda_{\rm ab}^{-2}(0)$ and is related to the apical and/or chain oxygen atoms.

Our results, namely a strong OIE stemming from the planar oxygen ions and a negligible OIE associated with the $c-$axis ions, suggest a polaronic nature of the superconducting carriers. This is consisitent with the basic idea of K. Alex M\"{u}ller, leading him and Georg Bednorz to the discovery of high-temperature superconductivity,\cite{Bednorz_Z-Phys-B_1986} that Jahn-Teller (JT) polarons provide a much stronger electron pairing glue than the conventional BCS electron-phonon interaction. Polarons are objects, which strongly couple to the lattice dynamics and renormalize those through an exponential reduction in the electronic kinetic energy.  The heavy mass of the polaron slows down the electron hopping and the OIE is a consequence of this renormalization process. Instead of a commonly inferred single band model, three bands need to be taken into account, namely a first nearest and second nearest neighbor hopping together with an inter-planar term.\cite{Bussmann-Holder_JSNM_2009} While the nearest neighbor interaction plays no role (or even reversed one) for the OIE, the second nearest neighbor term provides the correct trend of the doping dependent OIE. Real quantitative agreement is, however, obtained by including also the inter-planar hopping leading to the conclusion that JT polaron formation takes place and $c-$axis related hopping is polaronically renormalized. It is important to emphasize that coexisting order parameters, namely $d-$ and $s-$wave, need to be included in the analysis and that both are important to understand the experimental observations. Even though the $d-$wave component is dominantly present, modulations and stripe like features of the lattice together with phonon anomalies can only be understood by considering an additional $s-$wave order parameter.
In a detailed analysis of previous experimental results, coupled  gaps have been self-consistently calculated,\cite{Keller_MaterToday_2008} from which it is apparent that both components are needed to arrive at a consistent agreement with the experiments. \cite{Khasanov_La214_PRL_2007, Khasanov_Y123_PRL_2007,Khasanov_Y124_JSNM_2008}
Thus, we can close the circle of the discovery, its background together with K. Alex M\"{u}ller’s intuition by concluding that JT polarons play a decisive role for the pairing glue in HTSs.

\end{document}